\documentstyle[aps,prl,epsf,amsfonts]{revtex}
\newcommand{\eqb}{\begin{equation}}
\newcommand{\eqe}{\end{equation}}
\newcommand{\dmb}{\begin{displaymath}}
\newcommand{\dme}{\end{displaymath}}
\newcommand{\pd}{\partial}
\newcommand{\ep}{\varepsilon}
\newcommand{\eab}{\begin{eqnarray}}
\newcommand{\eae}{\end{eqnarray}}

\newcommand{\e}{\mbox{e}}

\newcommand{\nc}{\newcommand}
\nc{\lab}{\label}
\nc{\eq}{Eq.\,(\ref}
\nc{\eqs}{Eqs.\,(\ref}
\nc{\tm}{\tiny\mbox}
\nc{\vivi}{very interesting and very important}
\nc{\al}{\alpha}
\nc{\ga}{\gamma}
\nc{\de}{\delta}
\nc{\ze}{\zeta}
\nc{\et}{\eta}

\nc{\Th}{\Theta}
\nc{\ka}{\kappa}
\nc{\lam}{\lambda}
\nc{\rh}{\rho}
\nc{\si}{\sigma}
\nc{\ta}{\tau}
\nc{\up}{\upsilon}
\nc{\ph}{\phi}
\nc{\ch}{\chi}
\nc{\ps}{\psi}
\nc{\om}{\omega}
\nc{\Ga}{\Gamma}
\nc{\De}{\Delta}
\nc{\La}{\Lambda}
\nc{\Si}{\Sigma}
\nc{\Up}{\Upsilon}
\nc{\Ph}{\Phi}
\nc{\Ps}{\Psi}
\nc{\Om}{\Omega}
\nc{\ptl}{\partial}
\nc{\del}{\nabla}
\nc{\be}{\begin{equation}}
\nc{\ee}{\end{equation}}
\nc{\bea}{\begin{eqnarray}}
\nc{\eea}{\end{eqnarray}}
\nc{\ov}{\overline}
\nc{\gsl}{\!\not}

\draft

\begin{document}

\title{Deconfining by Winding: The Magnetic Monopole 
Vacua at High Temperatures}
\author{Ralf Hofmann}
\address{Theoretical Physics Institute,  
         University of Minnesota,  
         Minneapolis, MN 55455, USA }

\maketitle

\begin{flushright}
TPI-MINN-00/32  \\
June 2000
\end{flushright}

\begin{abstract}

Characterizing the vacuum 
of a thermalized SU(3) Yang-Mills theory in the dual 
Ginzburg-Landau description, the possibility of topologically 
nontrivial, classical monopole fields in the deconfining phase is 
explored. These fields are assumed to be 
Bogomol'nyi-Prasad-Sommerfield (BPS) 
saturated solutions along the compact, 
euclidean time dimension. A corresponding, gauge invariant 
monopole interaction is constructed. 
The model passes first tests. In particular, a 
reasonable value for the critical 
temperature is obtained, and the partial persistence of 
nonperturbative features in the deconfining 
phase of SU(3) Yang-Mills theory, as it is 
measured on the lattice, follows naturally.

\end{abstract} 



\section{Introduction}

The idea of explaining confinement in SU(N) Yang-Mills 
theory via the dual Meissner 
effect has been put forward by Mandelstam and 't Hooft in the 
late seventies and early eighties \cite{HM}. In this picture 
magnetic monopoles occur as point-like objects in three-space 
which are charged under the gauge group U(1)$^{\tiny\mbox{N}-1}$ 
constituting the maximal abelian subgroup 
to leave a certain gauge condition intact. The condition is that 
a certain matrix quantity (say the nonabelian field strength component ${\bf G}_{23}$), 
transforming {\em homogenously} under the gauge group, must be diagonal. It is shown 
that at points with a degeneracy of eigenvalues 
the transformation leading to the fulfillment of the 
gauge condition causes singularities of 
the colormagnetic flux, that is, magnetic monopoles. 
Fixing the gauge in a maximal abelian manner, one 
hence obtains an abelian theory of electric and magnetic charges. 
Working with N-1 (instead of 2(N-1)) 
abelian gauge fields only, the Lagrangian of the theory necessarily contains 
nonlocal terms which one would like to avoid \cite{Bard}. 
The way out is the introduction of a twin set of 
gauge fields \cite{Bard,Zwanziger}. The resulting theory contains 
electrically and magnetically charged matter 
fields and sets of (N-1) electric and magnetic 
abelian gauge fields $\vec {A}_\mu$ and $\vec {B}_\mu$, respectively. 
A {\em phenomenological} 
self-interaction of the magnetic monopole fields 
is assumed to yield a spontaneously generated condensate breaking the 
U(1)$^{\tm{N}-1}$ symmetry \cite{Suzuki}. As a result, the set of gauge fields $\vec {B}_\mu$, 
interacting with the magnetic charges \footnote{This sector is a dual Ginzburg-Landau theory.}, 
becomes massive and the colorelectric flux between two heavy 
quarks of opposite color charge, separated by a distance $R$, 
is confined to a straight tube (string) of constant tension $\sigma$ causing a potential 
linearily rising with $R$. This is the celebrated confinement mechanism. 
The occurence of a monopole 
condensate, starting from first principles, has never been 
proven analytically. However, a set of local and non-local lattice 
interactions of monopole currents has been fitted to the output 
of Monte-Carlo lattice simulations of maximal abelian gauge 
fixed SU(2) in Refs.\,\cite{Shiba,Cherno}. As a result the 
condensation of monopoles in the continuum limit 
was proven using energy-entropy balance arguments. An extension of 
this work to the SU(3) case was reported in Ref.\,\cite{Yama}.

It has been shown on the lattice in the more recent past that a 
thermalized SU(3) Yang-Mills theory undergoes a deconfinement phase 
transition at critical temperatures $T^{lat}_c\sim 200-300$ MeV 
(see for example Ref.\,\cite{Ha}). 
In the framework of the dual Ginzburg-Landau theory the deconfinement phase transition 
at finite temperature has been first discussed 
in Ref.\,\cite{Monden}. In Ref.\,\cite{TokiT} 
the critical temperature $T_c$ of the deconfinement phase transition 
was determined as the point where a thermal 
one-loop effective potential starts possessing  
a trivial, absolute minimum. The analysis was based on a 
Higgs-like (renormalizable) self-interaction 
of the monopole fields. From the $T$-dependence of the monopole condensate the 
temperature evolutions of the vector and monopole 
masses $m_B$, $m_\phi$ were calculated which in turn determine 
the thermal string tension $\sigma$. 

In the present work we start with a different philosophy. 
Based on the assumption of a gauge invariant potential for the 
(self-)interaction of {\em classical} monopole fields $\phi_k,\ k=1,2,3$, and 
considering only the U(1)$^{2}$ electric and magnetic 
gauge fields to fluctuate, we obtain a picture for the 
deconfinement phase transition and 
quantitative results for nonperturbative features in the deconfining phase. Thereby, 
the potential is constructed to allow for both, 
topologically trivial and nontrivial ($T>0$) BPS saturated solutions for each of 
the monopole fields $\phi_k$. As far as ground 
state properties are concerned, we feel that the assumption of 
classical monopole dynamics is justified by 
the observation that nonperturbative quantities 
like the string tension $\sigma$ could be 
determined reasonably well 
within the meanfield approximation in Refs.\,\cite{Suzuki,Toki0}. 

The starting point of 
our investigations is the dual 
Ginzburg-Landau theory given by the following 
Lagrangian \cite{Suzuki}
\eqb
\label{Lag}
{\cal L}_{DGL}=-\frac{1}{4}(\pd_\mu\vec{B}_\nu-\pd_\nu\vec{B}_\mu)^2+\sum_{k=1}^3\left\{\left|(i\pd_\mu-
g\vec{\ep}_k\cdot\vec{B}_\mu)\phi_k\right|^2-V_k(\phi_k,\bar{\phi}_k)\right\}\ .
\eqe
This theory decribes the magnetic sector emerging from SU(3) Yang-Mills theory 
due to maximal-abelian gauge fixing \cite{Suzuki}. 
In contrast to the electric, abelian gauge group the magnetic U(1)$^2$ 
(with gauge field $\vec{B}_\mu=(B^3_\mu,B^8_\mu)$) is believed to be 
spontaneously broken by nonvanishing VEVs of the monopole fields $\phi_k$.     
In Eq.\,(\ref{Lag}) $g\vec{\ep}_i$ denote the effective magnetic 
charges with $\vec{\ep}_1=(1,0),\ 
\vec{\ep}_2=(-1/2,-\sqrt{3}/2),\ \vec{\ep}_3=(-1/2,\sqrt{3}/2)$, and the fields $\phi_k$ 
satisfy the constraint \cite{Suzuki}
\eqb
\label{pc}
\sum_{k=1}^{3}\mbox{arg}\,\phi_k=0\ .
\eqe
The potential 
\eqb
V\equiv \sum_{k=1}^3V_k(\phi_k,\bar{\phi}_k) 
\eqe
is introduced {\em phenomenologically} to 
account for the self-interaction of the monopole 
fields $\phi_k$. For the sake of renormalizability and the desired 
feature of spontaneous breaking of the magnetic gauge symmetry 
it is usually chosen to be of the Higgs form 
\eqb
\label{pot}
\lambda\sum_{k=1}^3\left(\bar{\phi}_k\phi_k-v^2\right)^2\ .
\eqe
Corresponding to Eq.\,(\ref{pot}), the one-loop effective potential 
at finite temperature, calculated in 
euclidean space-time with compact time dimension of 
size $\beta\equiv 1/T$, possesses a global minimum of vanishing condensate 
for $T>T_c\sim 500$ MeV \cite{TokiT}. This result is probably due to the sole 
consideration of quadratic fluctuations, since there are indications that 
for example a quartic term is parametrically $not$ small (In one-loop approximation 
the coupling constant for this term is $\lambda\sim 25$!). In order to bring $T_c$ 
down to values compatible with lattice 
experiments ($T_c^{lat}\sim 200-300$ MeV) the authors of Ref.\,\cite{TokiT} 
had to introduce a $T$-dependence of $\lambda$ by hand.    

Here, we postulate that the occurence of a phase 
transition is due to the thermalized system 
admitting topologically nontrivial 
ground state solutions in the monopole sector. Accordingly, we 
introduce a gauge invariant (self-)interaction for the classical fields $\phi_k$. 
To construct the potential we furthermore assume that the 
relevant solutions to the corresponding field
equations are BPS saturated. This is in analogy to the case of instantons in 
non-gauge-fixed Yang-Mills theories, where the condition of self-duality is equivalent to the 
saturation of the Bogomoln'yi bound for the euclidean action \cite{Schaefer}. Since at $T=0$ the 
ground state solution is a constant and at $T>0$ changes are 
induced by the finite size of the 
euclidean, compact time dimension it is reasonable to assume that 
the topologically nontrivial structure of the vacuum 
above $T_c$ is brought about by a sole 
dependence on euclidean time $\tau$. In Refs.\,\cite{ShifDva,Losev} 
it was shown in the framework of supersymmetric theories that for a scalar sector 
to admit nontrivial BPS saturated solutions along a compact dimension the 
target space must possess noncontractable cycles. In Ref.\,\cite{Hof} a generalized 
Wess-Zumino model, possessing a discrete set of $N+1$ nonzero vacua 
(proportional to the $N+1$ unit roots), was investigated. These vacua are interpolated 
by BPS saturated, periodic solutions along the compact dimension. As in Ref.\,\cite{Hof}, 
we concentrate on the case, where there is one single pole at $\phi_k=0$ on the rhs of the BPS equation. 
Since the monopole sector of the theory should {\bf 1}) admit nonzero, constant 
solutions at $T=0$, {\bf 2}) be gauge invariant under the magnetic U(1)$^2$, and {\bf 3}) admit 
nontrivial, periodic, BPS saturated solutions for $T>0$ the potential is given as \cite{Hof}             
\eqb
\label{Veff}
V_k(\phi_k,\bar{\phi}_k)=\lim_{N\to\infty}\,\left\{\frac{\La^6}{\bar{\phi}_k\phi_k}+\ka^2\La^{-2(N-2)}
(\bar{\phi}_k\phi_k)^N-2\,\ka\Lambda^{5-N}\frac{1}{\bar{\phi}_k\phi_k}\mbox{Re}\,\phi_k^{N+1}\right\}\ . 
\eqe
Thereby, $\La$ is a mass-parameter, and $\ka$ is 
some dimensionless coupling constant. Considering the rhs of Eq.\,(\ref{Veff}) at finite $N$, the 
potential explicitely breaks the magnetic U(1)$^2$ 
gauge symmetry 
\eqb
\label{gauge-sy}
\vec{B}_\mu\,\to\, \vec{B}_\mu+1/g\,\pd_\mu\vec{\theta}(x)\ ,\ \ \ 
\phi_k\,\to\,\e^{-i\vec{\ep}_k\cdot\vec{\theta}(x)}\phi_k 
\eqe
down to Z$^2_{N+1}$ due to the term $\mbox{Re}\,\phi_k^{N+1}$. The 
vacua in this case are given as \cite{Hof}
\eqb
\frac{\La}{\ka^{1/(N+1)}}\,\left(\e^{\frac{2\pi\alpha_1}{N+1}i},\e^{\frac{2\pi\alpha_2}{N+1}i},
\e^{\frac{2\pi\alpha_3}{N+1}i}\right)\ ,\ \ \ (\alpha_k\in{\bf Z},\ -N\le\alpha_k\le N,\ 
\sum_{k=1}^3\alpha_k=0)\ .
\eqe
Only in the 
limit of large $N$ is the gauge symmetry restored. 
For the purpose of qualitative illustration Fig. 1 indicates 
the potential $V_k$ for $\ka=1$, $N=500$, and dimensionless $\La=0.5$. 
\begin{figure}
\vspace{6.2cm}
\includegraphics{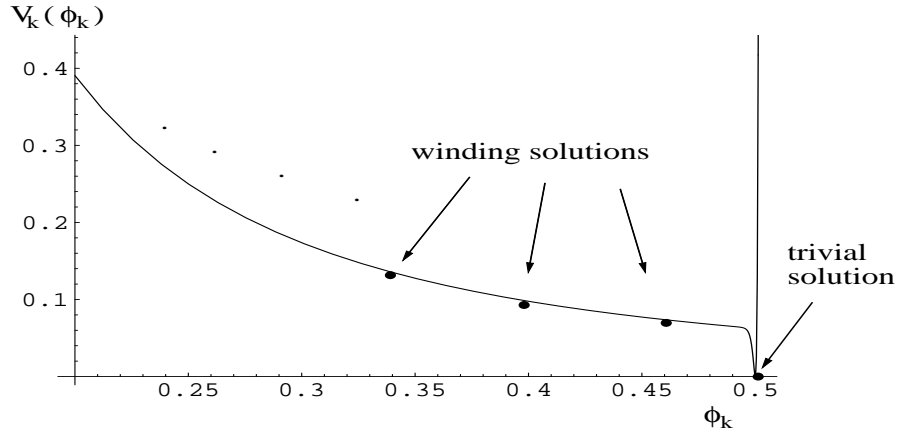}
\caption{The monopole potential $V_k$ as a 
function of real $\phi_k$ for $\ka=1,\,\La=0.5$, and $N=500$. Indicated are 
the potential energy densities of topologically trivial and 
nontrivial solutions to the BPS equations.} 
\label{} 
\end{figure}
The BPS equations, corresponding to the potential of Eq.\,(\ref{Veff}), read
\eqb
\label{BPS}
\pd_\tau\phi_k=\bar{V}^k_{1/2}(\bar{\phi_k})\ ,\ \ \pd_\tau\bar{\phi}_k=V^k_{1/2}(\phi_k)\ .
\eqe
The rhs are only fixed up to phase factors $\e^{i\delta}$, $\e^{-i\delta}$, 
respectively. From Eq.\,(\ref{Veff}) we have for {\em periodic} solutions 
\eqb
\label{V1/2}
V^k_{1/2}(\phi_k)=\left\{\begin{array}{c} \pm i\frac{\La^3}{\phi_k}\ ,\ \ \ \  (|\phi_k|<\La)\\ 
0\ ,\ \ \ \ \ \ \ \ \ (|\phi_k|=\La)\\ 
\ \infty\ ,\ \ \ \ \ \ \ (|\phi_k|>\La)\end{array}\right.\ .
\eqe
The paper can be outlined as follows: In the next section we briefly review what has to be 
known about BPS saturated winding solutions along a compact 
dimension \cite{ShifDva,Losev,Hof}. We device a (not unique) label 
for the winding vacua, and we write down the gauge function to transform 
a winding vacuum to unitary gauge. Section 3 investigates the transition to the deconfined phase. 
The critical temperature is estimated by assuming thermal equilibrium between 
the winding vacuum and an ideal gluon
gas. The consequences of winding vacua in the deconfined phase are examined in 
Section 4. The temperature evolutions of the monopole condensate 
(and thereby that of $m_B$) and the string tension are elucidated. Section 5 summarizes the results, 
comments on the approximations made, 
and gives the conclusions.

\section{BPS saturated winding solutions}

For the case $|\phi_k|<\La$ periodic solutions ($\phi_k(0)=\phi_k(\beta)$) 
to Eqs.\,(\ref{BPS}) subject to Eq.\,(\ref{V1/2}) have been discussed 
in Refs.\,\cite{ShifDva,Losev} within the framework of supersymmetric theories. For a compact, 
euclidean time dimension of length $\beta\equiv 1/T$ 
the set of topologically distinct solutions is
\eqb
\label{wind}
\phi^{n_k}_k(\tau)=\sqrt{\frac{\La^3\beta}{2|n_k|\pi}}\,
\e^{2n_k\pi i\frac{\tau}{\beta}}\ ,\ \ (n_k\in{\bf Z})\ .
\eqe
Thereby, the sign of $n_k$ corresponds to the choice of 
phase in the BPS equations. 
Due to the phase constraint of Eq.\,(\ref{pc}) the sets of 
solutions $(\phi^{n_1}_1,\phi^{n_2}_2,\phi^{n_3}_3)$ can be labeled by 
the integers $n_1$ and 
$m$, which are both odd {\em or} even. The winding numbers 
$n_2,\,n_3$ are then given as $-n_1/2\mp m/2$, respectively. 
With the gauge transformation of Eq.\,(\ref{gauge-sy}) the function 
$\vec{\theta}_{n_1,m}(\tau)$, transforming to the unitary gauge 
Re\,$\phi_k(\tau)>0$, Im\,$\phi_k(\tau)=0$, reads
\eqb
\label{uni}
\vec{\theta}_{n_1,m}(\tau)=\frac{2\pi}{\beta}\,
\tau\left(\begin{array}{c}n_1\\ \frac{m}{\sqrt{3}}\end{array}\right)\ .
\eqe
Note that this nonperiodic function 
leaves the periodicity of the gauge field $\vec{B}_\mu$ intact.

\section{Deconfinement phase transition}

Eqs.\,(\ref{BPS}) admit the solution $\phi^0_k\equiv\La$ with winding number 
and energy density zero, and the winding solutions of the previous section. 
We identify the set $(\phi^0_1,\phi^0_2,\phi^0_3)$ 
with the confining vacuum at low temperatures. In this regime 
the spectrum consists of glueballs with masses that are 
much larger than the prevailing 
temperatures \cite{Toki0}. The thermal equilibrium between the 
vacuum "medium" and its excitations is essentially 
realized at pressure zero, which indeed is 
observed on the lattice \cite{Beinl}. Above the deconfinement 
phase transition the ground state "medium" readily emits and absorbs 
(almost) free gluons under the 
influence of the heat bath like a black body emits and 
absorbs photons. We identify the genuine winding set 
of lowest potential energy density\footnote{By "genuine" we mean that 
each of the solutions $\phi_k$ is winding. 
Note, however, that 
there are also semi-winding sets, for example $(n_1,m)=(1,1)$, 
which contain two winding fields 
and one field of winding number zero. Since none of the 
monopole fields $\phi_k$ should be singled out it is natural
to assume that either all $\phi_k$ are winding or none at all.}, 
represented by 
$(n_1,m)=(2,0)$, with the ground 
state just above the deconfinement transition. 
In this picture the confining vacuum is a perfect thermal insulator 
up to the transition, 
where its structure drastically changes by the absorption 
of an amount of latent heat per volume \cite{Beinl} 
equal to the gap $\Delta\ep^{(2,0)}$ between the potential $V$ of the 
zero winding and the lowest genuine winding set. 
We estimate the critical temperature $T_c$ 
of this transition by assuming 
the deconfining vacuum to behave like an 
incompressible, static fluid with traceless energy-momentum-tensor \cite{Cley} 
which is in thermal equilibrium with an ideal gas of gluons. In this case we 
have the equation of state  
\eqb
\ep=3p
\eqe
for pressure $p$ and energy density $\ep$ 
for both the vacuum (vac) and the gluon gas (gg). The equilibrium 
condition of equal pressures, 
\eqb
\label{peq}
p_{vac}=p_{gg}\ , 
\eqe
then takes the following form \cite{Cley}
\eqb
\label{eq} 
\Delta\ep^{(2,0)}=\frac{8}{15}\pi^2 T_c^4\ , 
\eqe
where 
\eqb
\label{del}
\Delta\ep^{(2,0)}=2\times\frac{\La^6}{\bar{\phi}^1\phi^1}+\frac{\La^6}{\bar{\phi}^2\phi^2}=8\pi\La^3T_c\ .
\eqe
Using Eqs.\,(\ref{eq}),(\ref{del}), we obtain
\eqb
\label{Tc}
T_c=\left(\frac{15}{\pi}\right)^{\frac{1}{3}}\La\sim 1.68\,\La\ .
\eqe
Adopting the value $\La=0.126$ GeV for the monopole 
condensate at $T=0$ from Refs.\,\cite{Toki0,TokiT}, Eq.\,(\ref{Tc}) 
yields $T_c=0.212$ GeV. This is compatible with the 
lattice results of $T_c=0.2-0.3$ GeV.

\section{The deconfining phase}
For a set of solutions with  
\eqb
l\equiv\sum_{k=1}^3|n_k|\ ,
\eqe
the equilibrium condition of Eq.\,(\ref{peq}) yields 
the following dependence of the 
temperature at equilibrium $T_l$ on $l$
\eqb
\label{Tl}
T_l=\left(\frac{15l}{4\pi}\right)^{\frac{1}{3}}\La\ .
\eqe
Modulo degeneracy the number of 
equilibrium states per temperature interval $\rho(T)$ thus grows quadratically
with increasing $T_l$ 
\eqb
\rho(T)\sim\frac{dl}{dT_l}=\frac{4\pi}{5}\frac{T_l^2}{\La^3}\ .
\eqe
For temperatures $T\not=T_l$ the system is thermodynamically 
unstable within the above approximations, and hence 
it must evolve to the nearest equilibrium point. 
Let us now look at the change of the average monopole condensate 
\eqb
\phi^{(n_1,m)}_{av}\equiv\frac{1}{3}\sum_{k=1}^3 \phi^{n_k}_k
\eqe
across the transition from the confining to the deconfining phases. 
Transforming the winding solution of Eq.\,(\ref{wind}) to 
the unitary gauge by means of Eqs.\,(\ref{gauge-sy}),(\ref{uni}), 
we obtain the following expression 
for the individual condensate, corresponding to winding $n_k$ and 
contributing to an $l$-winding vacuum   
\eqb 
\phi_k^{n_k}=\frac{\La^{3/2}}{\sqrt{2\pi |n_k| T_l}}=\frac{\La}{\sqrt{(30\pi^2 l)^{1/3}|n_k|}}\ .
\eqe
For the configuration $(n_1,m)=(2,0)$ we obtain 
\eqb 
\phi^{(2,0)}_{av}\sim 0.28\,\La\ .
\eqe
In Fig. 2 the evolution of the average monopole 
condensate is depicted up to $T\sim0.41$ GeV. The arrows point at states of thermal equilibrium. 
The corresponding representatives, labeled by $(n_1,m)$, evolve into 
nonequilibrium until the next equilibrium state becomes available. 
\begin{figure}
\vspace{7.0cm}
\includegraphics{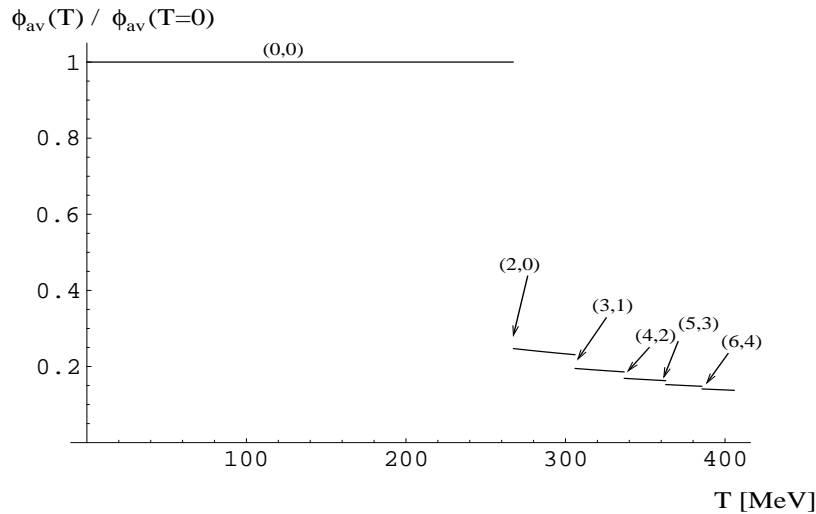}
\caption{The average monopole field $\phi_{av}(T)$ in units of 
$\phi_{av}(T=0)$. For further explanation see text.} 
\label{} 
\end{figure}
In unitary gauge the monopole fields are real, and 
the spontaneous breakdown of the U(1)$^2$ magnetic gauge symmetry then transparently 
generates a mass $m_B$ for the field $\vec{B}_\mu$ which scales linearly 
with the average monopole condensate:
\eqb
m_B^{(n_1,m)}=\sqrt{3}\,g\,\phi^{(n_1,m)}_{av}\ .
\eqe
In Refs.\,\cite{Toki0,TokiT} the value $m_B\sim 0.5$ GeV was 
determined at $T=0$, yielding a magnetic gauge coupling of $g=2.3$.
   
How does the string tension $\sigma$ evolve? An analytical 
expression for $\sigma$ in terms of $g,\,m_B$, and $m_\phi$ 
was derived in Ref.\,\cite{Toki0} by calculating the potential between 
two heavy and static quarks of opposite color charge, separated by a distance $R$. 
Thereby, an effective Lagrangian 
for the dynamics of the electric $U(1)^2$ gauge field $\vec{A}_\mu$ 
is obtained by integrating out the massive, magnetic vector field 
$\vec{B}_\mu$ and by assuming that the monopole fields do 
not fluctuate about their VEVs. Applying convenient gauge fixing terms and 
introducing external, static color currents, which couple to $\vec{A}_0$, one obtains an 
effective action. From this action one extracts the potential energy $V_{\bar{q}q}(R)$ 
of the static sources. It 
assumes the following form 
\eqb
\label{potR}
V_{\bar{q}q}(R)=-\frac{\vec{Q}^2}{4\pi}\left(\frac{\e^{-m_B^2 R}}{R}+
\frac{m_B^2}{2}\ln\left[1+\frac{m_\phi^2}{m_B^2}\right]\,R\right)\ ,
\eqe
where $\vec{Q}=Q_3^2+Q_8^2=e^3/3$ \cite{Suzuki}. From Eq.\,(\ref{potR}) 
$\sigma$ can be read off as
\eqb
\label{sig}
\sigma=\frac{2\pi m_B^2}{3g^2}\ln\left[1+\frac{m_\phi^2}{m_B^2}\right]\ .
\eqe
Thereby, we have used the relation $e\cdot g=4\pi$ linking 
the magnetic with the electric charge. Since {\em static} test charges 
were assumed finite temperature does not necessitate any new considerations, and 
thus $\sigma$ is still given by Eq.\,(\ref{sig}) 
with an implicit $T$-dependence via $m_B(T)$ and $m_\phi(T)$. 
\begin{figure}
\vspace{6.8cm}
\includegraphics{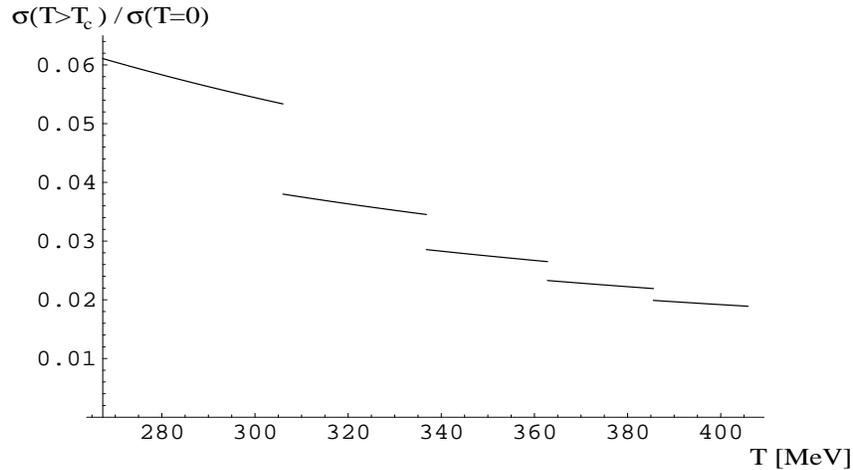}
\caption{The string tension $\sigma(T>T_c)$ in units of $\sigma(T=0)$ in the deconfining phase.} 
\label{} 
\end{figure}
The appearance of the 
monopole mass $m_\phi>m_B$ under the logarithm in Eq.\,(\ref{sig}) 
is due to its role as an ultraviolet 
cutoff for the integration over transverse momenta in the expression for 
the linear part of the potential in Eq.\,(\ref{potR}). 
This is justified by the observation that $m_B\sim0$ inside the 
flux tube of radius $\rho\sim m_\phi^{-1}$ in a type II dual superconductor 
(see Ref.\,\cite{TokiT} and Refs. therein). Since we work with classical fields $\phi_k$ we cannot 
asign a mass to the fluctuations of the monopole fields about the 
background of a classical ground state. We may, however, assume that $\rho^{-1}$ 
scales with temperature in the same way as $m_B$ does. Since the dependence of $\sigma$ 
on the cutoff is logarithmic the result is not dramatically 
sensitive to the exact $T$ dependence of 
$\rho^{-1}$, which we will make explicit by distinguishing the following cases:\\  
(a) $\rho^{-1}(0)=\rho^{-1}(T)=1.26$ GeV \cite{Toki0} $\Rightarrow$ 
\eqb
\frac{\sigma(T_c)}{\sigma(0)}\sim (0.28)^2\frac{\ln(1+(1.26)^2/(0.5\times0.28)^2)}
{\ln(1+(1.26)^2/(0.5)^2)}\sim 0.19\ .
\eqe
(b) $\rho^{-1}(T)\propto m_B(T)$ $\Rightarrow$ 
\eqb
\frac{\sigma(T_c)}{\sigma(0)}\sim (0.28)^2\sim 0.08\ .
\eqe
Hence, the difference roughly amounts to a factor 2. Assuming case (b), we show the $T$-dependence 
of $\sigma(T)/\sigma(0)$ for the deconfining phase in Fig. 3.

\section{Summary and discussion}

We considered a thermalized dual Ginzburg-Landau theory modelling hot SU(3) Yang-Mills theory. 
From the postulate that in contrast to the confining phase 
the ground state of the deconfining phase is characterized by 
topologically nontrivial, BPS saturated solutions to the 
classical equations of motion of the monopole sector we deviced the corresponding, 
gauge invariant interaction. As a consequence, the average monopole 
field undergoes a drastic decrease to about 1/4 of its 
zero temperature value across the phase boundary. Scaling linearly 
with the monopole condensate, 
the same applies to the mass of the magnetic vector fields generated by the 
spontaneous breaking of the magnetic U(1)$^2$ gauge symmetry. Working with the zero 
temperature value for the monopole condensate of Refs.\,\cite{Toki0,TokiT}, 
we obtain a rather reasonable critical temperature $T_c\sim 212$ MeV in 
contrast to the effective potential calculation 
of Ref.\,\cite{TokiT}, where $T_c\sim 500$ MeV, 
and the more realistic value of $T_c\sim200$ MeV was obtained by 
introducing an ad hoc $T$-dependence of the dimensionless 
coupling constant $\lambda$. 

Depending on the assumption about 
the $T$-dependence of the flux tube radius $\rho$ the string tension $\sigma$ decreases 
to about 1/12..1/5 of its value at $T=0$ across the phase boundary which is 
compatible with lattice measurements \cite{Beinl}. To determine $T_c$ 
we assumed a {\em free} gluon gas which is a rough approximation due to the fact 
that nonperturbative features do persist to a 
sizable extend in the deconfined phase. This is indicated 
by nonvanishing values of the monopole condensate 
$\phi_{av}$ and the string tension 
$\sigma$. At higher temperatures the ideal gas approximation should become better, 
and the model predicts a slow decrease of these quantities. 
For example, at $T\sim1.5\, T_c$
we have 
\eqb
\frac{\phi_{av}(T=1.5\, T_c)}{\phi_{av}(T=0)}\sim 
\frac{1}{2}\times\frac{\phi_{av}(T=T_c)}{\phi_{av}(T=0)}\ .
\eqe
On the lattice and in effective potential calculations the deconfinement 
phase transition has been determined 
to be of first order \cite{TokiT,Beinl,Karsch}. In contrast, SU(2) Yang-Mills theory 
exhibits a second order phase transition \cite{Karsch}. The 
model of the present work can be easily adapted to 
the dual Ginzburg-Landau  
theory describing maximal abelian 
gauge fixed SU(2) Yang-Mills theory. The qualitative results 
would then be the same. Hence, the model is too coarse 
to resolve the order of the phase transition. 
However, it gives a reason for the residual, nontrivial structure 
of monopole condensation in SU(2) and SU(3) Yang-Mills theories at high temperatures, it predicts a 
reasonable value of the critical temperature, and it 
incorporates the low temperature limit in the form 
of a constant solution to the BPS equations. The existence of a nontrivial structure 
of monopole condensation above the phase transition has been pointed out in Ref.\,\cite{Lau} 
a long time ago based on lattice investigations.  

To conclude, we showed that in a thermalized, generalized dual 
Ginzburg-Landau theory, 
describing the magnetic part of SU(3) Yang-Mills 
theory subject to maximal abelian gauge fixing, the postulate of nontrivial, BPS saturated 
ground state solutions of the magnetic monopole sector 
in the deconfining phase leads to acceptable results. In particular, 
the partial conservation of nonperturbative features in the high temperature 
regime is a natural consequence in this picture.

\section*{Acknowledgments}

The author would like to thank V. Eletsky and M. Pospelov for 
stimulating discussions at an early stage of this
work. Continuous, useful conversations 
with T. ter Veldhuis are gratefully acknowledged. 
This work was funded by a fellowship of Deutscher 
Akademischer Austauschdienst (DAAD).

\bibliographystyle{prsty}

\begin{thebibliography}{10}


\bibitem{HM}
S. Mandelstam, Phys. Rep. C{\bf 23}, 245 (1976).
G. 't Hooft, Nucl. Phys. B{\bf 190}, 455 (1981). 

\bibitem{Bard}
K. Bardakci and S. Samuel, Phys. Rev. D{\bf 18}, 2849 (1978). 

\bibitem{Zwanziger}
D. Zwanziger, Phys. Rev. D{\bf 3}, 880 (1970).

\bibitem{Suzuki}
T. Suzuki, Prog. Theor. Phys. {\bf 80}, 929 (1988); {\bf 81}, 752 (1989).\\ 
S. Maedan and T. Suzuki, Prog. Theor. Phys. {\bf 81}, 229 (1989).

\bibitem{Shiba}
H. Shiba and T. Suzuki, Phys. Lett. B{\bf 351}, 519 (1995).

\bibitem{Yama}
K. Yamagishi, S. Kitahara and T. Suzuki, JHEP {\bf 0002}, 012 (2000).

\bibitem{Cherno}
M. N. Chernodub, S. Fujimoto, S. Kato, M. Murata, 
M. I. Polikarpov and T. Suzuki, hep-lat/0006025, to appear in Phys. Rev. D.

\bibitem{Toki0}
H. Suganuma, S. Sasaki and H. Toki, Nucl. Phys. B{\bf 435}, 207 (1995).

\bibitem{Monden}
H. Monden, T. Suzuki and Y. Matsubara, Phys. Lett. B{\bf 294}, 100 (1992)

\bibitem{TokiT}
H. Ichie, H. Suganuma and H. Toki, Phys. Rev. D{\bf 52}, 2944 (1995).

\bibitem{Ha}
T. Hashimoto {\em et al.}, Phys. Rev. D{\bf 42}, 620 (1990).

\bibitem{ShifDva}
G. Dvali and M. Shifman, Phys. Lett. B{\bf 454}, 277 (1999).

\bibitem{Losev}
X. Hou, A. Losev and M. Shifman, Phys. Rev. D{\bf 61}, 085005 (2000). 

\bibitem{Schaefer}
T. Schafer and E. V. Shuryak, Rev. Mod. Phys. {\bf 70}, 323 (1998).

\bibitem{Hof} 
R. Hofmann, hep-th/0004178, to appear in Phys. Rev. D. 

\bibitem{Cley}
J. Cleymans, E. Nyk\"anen and E. Suhonen, Phys. Rev. D{\bf 33}, 2585 (1986). 

\bibitem{Beinl}
B. Beinlich {\em et al.}, Eur. Phys. J. C{\bf 6}, 133 (1999).

\bibitem{Karsch}
F. Karsch, "THE DECONFINEMENT TRANSITION IN FINITE TEMPERATURE LATTICE 
GAUGE THEORY", Invited talk given at the 'Enrico Fermi' Int. School of 
Physics, Varenna, Italy, Jun 26-Jul 6, 1984.

\bibitem{Lau}
M. L. Laursen and G. Schierholz, Z. Phys. C{\bf 38}, 501 (1988).

\end{thebibliography}

\end{document}